\documentclass[conference]{IEEEtran}
\usepackage{cite}
\usepackage[ruled,vlined]{algorithm2e}
\usepackage{amsmath,amssymb,amsfonts}
\usepackage{graphicx}
\usepackage{textcomp}
\usepackage{xcolor}
\usepackage{multirow}
\usepackage{multicol}
\usepackage{siunitx}
\usepackage{makecell}
\usepackage[a4paper, total={184mm,239mm}]{geometry}
\def\BibTeX{{\rm B\kern-.05em{\sc i\kern-.025em b}\kern-.08em
    T\kern-.1667em\lower.7ex\hbox{E}\kern-.125emX}}

\usepackage{xcolor}

\usepackage{graphicx}

\usepackage{tikz}
\usepackage[most]{tcolorbox}
\usepackage{gensymb}
\usepackage{todonotes}
\usepackage{amsmath,amsfonts}
\usepackage[noend]{algpseudocode}
\usepackage[symbol]{footmisc}

\usepackage{tikz}
\usepackage{amsmath}
\usepackage{multirow}
\usepackage{booktabs}

\newcommand*\circled[1]{\tikz[baseline=(char.base)]{
            \node[shape=circle,fill,inner sep=1pt] (char) {\textcolor{white}{#1}};}}

\usepackage{xcolor}
\usepackage{pgfplots}
\usepackage{tikz}
\pgfplotsset{compat=1.8}
\pgfplotsset{
    width=\textwidth,
}
\usepackage{lipsum}
\usepackage{pgfplotstable}
\usetikzlibrary{pgfplots.groupplots}

\usepackage{tikz}

\definecolor{codegreen}{rgb}{0,0.6,0}
\definecolor{codegray}{rgb}{0.5,0.5,0.5}
\definecolor{codepurple}{rgb}{0.58,0,0.82}
\definecolor{mGreen}{rgb}{0,0.6,0}
\definecolor{mGray}{rgb}{0.5,0.5,0.5}
\definecolor{mPurple}{rgb}{0.58,0,0.82}
\definecolor{backcolour}{rgb}{0.95,0.95,0.92}

\definecolor{RYB1}{RGB}{80, 99, 42}
\definecolor{RYB2}{RGB}{215, 227, 191}
\definecolor{RYB3}{RGB}{198, 187, 174}
\definecolor{RYB4}{RGB}{146, 205, 220}
\definecolor{RYB5}{RGB}{238, 144, 34}
\definecolor{RYB6}{RGB}{142, 172, 59}

\pgfplotscreateplotcyclelist{colorbrewer-RYB}{
{RYB1!50!black,fill=RYB1},
{RYB2!50!black,fill=RYB2},
{RYB3!50!black,fill=RYB3},
{RYB4!50!black,fill=RYB4},
{RYB5!50!black,fill=RYB5},
{RYB6!50!black,fill=RYB6},
}

\usepackage{lscape} 
\usepackage{pdflscape}
\usepackage{afterpage}
\usepackage{capt-of}
\usepackage{listings}
\usepackage{float}
\usepackage{array,multirow,graphicx}
\usepackage{caption}
\usepackage{subcaption}
\usepackage{soul}
\usepackage{pifont}

\definecolor{ggreen}{HTML}{2CC225}
\definecolor{yyellow}{HTML}{C2C80A}
\definecolor{bbrown}{HTML}{8e4603}

\definecolor{codegreen}{rgb}{0,0.6,0}
\definecolor{codegray}{rgb}{0.5,0.5,0.5}
\definecolor{codepurple}{rgb}{0.58,0,0.82}
\definecolor{mGreen}{rgb}{0,0.6,0}
\definecolor{mGray}{rgb}{0.5,0.5,0.5}
\definecolor{mPurple}{rgb}{0.58,0,0.82}
\definecolor{backcolour}{rgb}{0.95,0.95,0.92}

\lstdefinestyle{CStyle}{
    commentstyle=\color{mGreen},
    keywordstyle=\color{magenta},
    numberstyle=\tiny\color{mGray},
    stringstyle=\color{mPurple},
    basicstyle=\sffamily\footnotesize,
    frame=lrtb,
    breakatwhitespace=false,         
    breaklines=true,                 
    captionpos=b,                    
    keepspaces=true,                 
    numbers=left,                    
    numbersep=5pt,                  
    showspaces=false,                
    showstringspaces=false,
    showtabs=false,                  
    tabsize=2,
    language=C
}

\lstdefinestyle{CStyle1}{
    commentstyle=\color{mGreen},
    keywordstyle=\color{magenta},
    numberstyle=\tiny\color{mGray},
    stringstyle=\color{mPurple},
    basicstyle=\sffamily\footnotesize,    frame=lrtb,
    breakatwhitespace=false,         
    breaklines=true,                 
    captionpos=b,                    
    keepspaces=true,                 
    numbers=left,                    
    numbersep=5pt,                  
    showspaces=false,                
    showstringspaces=false,
    showtabs=false,                  
    tabsize=2,
    language=C
}

\lstdefinestyle{mystyle}{
    commentstyle=\color{codegreen},
    keywordstyle=\color{magenta},
    numberstyle=\tiny\color{codegray},
    stringstyle=\color{codepurple},
    basicstyle=\sffamily\footnotesize,
    breakatwhitespace=false,         
    breaklines=true,                 
    captionpos=b,                    
    keepspaces=true,                 
    numbers=left,                    
    numbersep=5pt,                  
    showspaces=false,                
    showstringspaces=false,
    showtabs=false,                  
    tabsize=2,
    language=C
}
\lstdefinestyle{trans}{
    commentstyle=\color{codegray},
    numberstyle=\tiny\color{codegray},
    stringstyle=\color{codepurple},
     basicstyle=\sffamily\footnotesize,
    frame=lrtb,
    breakatwhitespace=false,         
    breaklines=true,                 
    captionpos=b,                    
    keepspaces=true,                 
    numbers=left,                    
    numbersep=5pt,                  
    showspaces=false,                
    showstringspaces=false,
    showtabs=false,                  
    tabsize=2,
     language=[x86masm]Assembler,  escapeinside={\%*}{*)},   
     }     

\ifCLASSINFOpdf
\else
\fi
\begin{document}
\pagestyle{plain}
%
\title{Logical Maneuvers: Detecting and Mitigating Adversarial Hardware Faults in Space}


\author{\IEEEauthorblockN{Fatemeh Khojasteh Dana\IEEEauthorrefmark{1}}
	\IEEEauthorblockA{Worcester Polytechnic Institute\\
    Worcester, MA\\
		fdana@wpi.edu}
	\and
	\IEEEauthorblockN{Saleh Khalaj Monfared\IEEEauthorrefmark{1}} 
	\IEEEauthorblockA{Worcester Polytechnic Institute\\
    Worcester, MA\\
		skmonfared@wpi.edu
        }
        \\ {\fontsize{7.5}{7.5}\selectfont \IEEEauthorrefmark{1} These authors contributed equally to this work}
	\and
	\IEEEauthorblockN{Shahin Tajik}
	\IEEEauthorblockA{Worcester Polytechnic Institute\\
    Worcester, MA\\
		stajik@wpi.edu}
        }


%

\IEEEoverridecommandlockouts
\makeatletter\def\@IEEEpubidpullup{5.5\baselineskip}\makeatother
\IEEEpubid{\parbox{\columnwidth}{
		{\fontsize{7.5}{7.5}\selectfont Workshop on Security of Space and Satellite Systems (SpaceSec) 2025 \\
			24 February 2025, San Diego, CA, USA \\
			ISBN 979-8-9919276-1-1 \\
            https://dx.doi.org/10.14722/spacesec.2025.23xxx  \\
			www.ndss-symposium.org 
            }
}
\hspace{\columnsep}\makebox[\columnwidth]{}}

\maketitle
\begin{abstract}
Satellites are highly vulnerable to adversarial glitches or high-energy radiation in space, which could cause faults on the onboard computer.
Various radiation- and fault-tolerant methods, such as error correction codes (ECC) and redundancy-based approaches, have been explored over the last decades to mitigate temporary soft errors on software and hardware.
However, conventional ECC methods fail to deal with hard errors or permanent faults in the hardware components.
This work introduces a detection- and response-based countermeasure to deal with partially damaged processor chips.
It recovers the processor chip from permanent faults and enables \emph{continuous operation with available undamaged resources} on the chip.
We incorporate digitally-compatible delay-based sensors on the target processor's chip to reliably detect the incoming radiation or glitching attempts on the physical fabric of the chip, even before a fault occurs.
Upon detecting a fault in one or more components of the processor's arithmetic logic unit (ALU), our countermeasure employs adaptive software recompilations to resynthesize and substitute the affected instructions with instructions of still functioning components to accomplish the task. 
Furthermore, if the fault is more widespread and prevents the correct operation of the entire processor, our approach deploys adaptive hardware partial reconfigurations to replace and reroute the failed components to undamaged locations of the chip.
To validate our claims, we deploy a high-energy near-infrared (NIR) laser beam on a RISC-V processor implemented on a 28~nm FPGA to emulate radiation and even hard errors by partially damaging the FPGA fabric.
We demonstrate that our sensor can confidently detect the radiation and trigger the processor testing and fault recovery mechanisms.
Finally, we discuss the overhead imposed by our countermeasure.

\end{abstract}


\section{Introduction}\label{sec:intro}
\begin{figure}[t]
\centering
\includegraphics[width=\columnwidth]{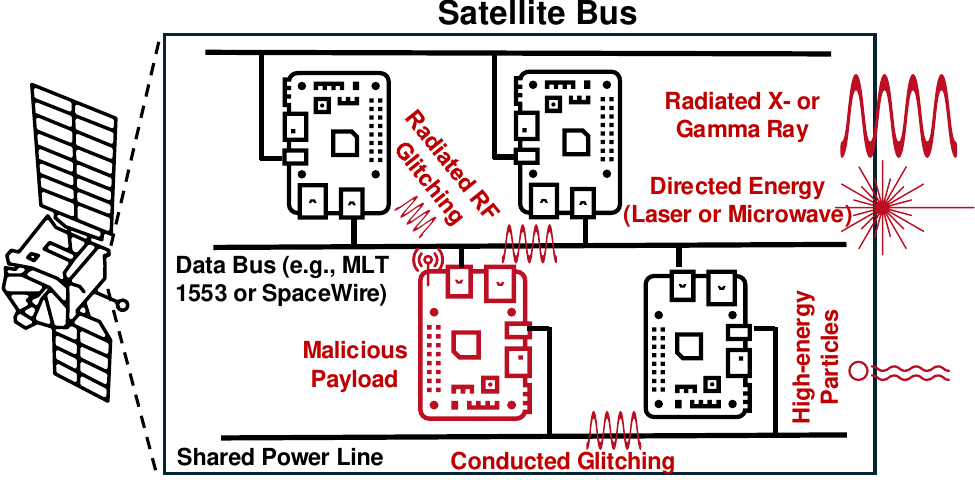}
\caption{Fault and radiation threats for hosted payloads on satellites}
\label{fig:threat}
\end{figure}

Satellites are vulnerable to cyberattacks that can destroy them without creating debris, minimizing collateral damage, or taking control of them entirely.
Attackers could manipulate satellites in various ways, such as depleting fuel, altering orbits, holding onboard computers for ransom, draining power, or damaging sensors by exposing them to direct sunlight~\cite{bailey2021cybersecurity}. 
The shift toward hosting multiple payloads on commercial satellite buses with shared resources (e.g., power source, transponders, data buses, etc.) creates novel cyber-physical vulnerabilities for hosted payloads if one of the payloads is infected and goes rogue, see Fig~\ref{fig:threat}.
Such a malicious payload can inject physical faults into the adjacent payloads' computers through the shared power lines, data buses, or simply vacuum using conducted or radiated electromagnetic glitches. 
While some proprietary commercial solutions exist for power isolation of hosted payloads to reduce the information leakage~\cite{l3harris}, it is unclear how effective they are against fault injection attacks. 
On the other hand, satellite onboard computers are threatened by external high-energy electromagnetic (EM) signals and charged particles caused by directed energy weapons, nuclear detonations, or cosmic rays in space~\cite{warning}.
Such high-energy EM pulses and particles can permanently damage the onboard computer chips.

Radiation hardening (RadHard) and fault-tolerant solutions have been widely researched in the literature to deal with such threats.
Error correction codes (ECCs) and redundancy-based methods, such as triple modular redundancy (TMR), are prime examples of detection and recovery schemes from faults.
While ECCs are effective against temporary soft errors or transient glitches to detect and correct the corrupted instruction or data, they are ineffective against permanent hard errors on the functional part of the chip as such hardware faults cannot be fixed by the running software.
Moreover, while TMR methods can deal with hard errors to some extent using voting, they impose very large overheads on the system, which have constrained size, weight, and power (SWaP) requirements.
In addition, due to the lack of sensing capability of these algorithmic countermeasures about the events occurring at the physical layer of a computing chip, they should always be active to respond to possible faults.
Interestingly, hard errors are not special to space applications as 
the shrinking trend of transistor technology leads to less reliable hardware~\cite{markoff2022tiny}. 
There are multiple reports that processors fail randomly due to hard errors in large data centers~\cite{dixit2021silent,hochschild2021cores}.
While it is straightforward to replace chips on the ground, it is virtually impossible to replace them in a spacecraft in orbit in most cases.

The physical detection of glitching attempts before fault occurrence and the response to it have been researched in the field of hardware security.
Deploying delay-based sensors is the primary way for sensing analog disturbances on the chip power delivery network (PDN) caused by voltage~\cite{moini2023fault}, EM~\cite{paquette2021visualizing}, and laser glitches~\cite{monfared2024laserescape}.
Upon detection of such disturbances, the chips can respond by locking the computation~\cite{yuce2018secure} or ignoring the processed data~\cite{moini2023fault} for a few prior clock cycles.
In contrast to classical hardware security scenarios, where the goal of a fault adversary is to break the confidentiality or violate the integrity of the system, the main goal for fault injection on space assets is to cause safety and availability issues~\cite{jerosecuring}.
Hence, while current detection methods could benefit the satellite onboard computers, most conventional response methods are unacceptable as they could make the system unavailable, creating safety issues for a satellite in orbit.

Hence, in this work, we ask the following research questions: \textit{
Is it possible to deploy sensors to detect the glitching/radiation attempt in space, recover from a hardware fault, and continue the computation even with a partially damaged processor circuit without imposing a large SWaP overhead and without having redundant processors on standby?}

\textbf{Our Contribution.} In this work, we positively answer the research questions above.
We first demonstrate that any glitching attempt or radiation causes voltage fluctuations on the PDN of the processor chip and, consequently, affects the propagation delays of signals (e.g., clock signal) on the chip.  
We show that such delay variations can be measured reliably by time-to-digital converters (TDCs)~\cite{alam2019ram,moini2023fault}.
Upon detection of delay variations, a software test is triggered to detect whether a fault has occurred.
By detecting a fault in the processor's arithmetic logic unit (ALU), software recompilation and resynthesis will be performed to replace the affected arithmetic or logic instruction with other available and still functioning instructions to accomplish the task.
On the other hand, if a fault makes the entire processor dysfunctional, adaptive hardware partial reconfigurations will be executed to move the damaged parts of the processor to a functional region on the chip.

To show the effectiveness of our approach, we realize a soft RISC-V processor on a common field programmable gate array (FPGA) manufactured with 28 nm technology with TDC sensors.
We emulate the radiation on the sensors by employing a near-infrared (NIR) laser beam. 
To emulate hard errors, we take a step further and create permanent damage on the configuration logic blocks of the FPGA fabric using our laser with higher power.
We demonstrate that our software and hardware recovery methods heal the processor and enable continuous operation without any interruption.
Finally, we discuss the overhead of our countermeasure in terms of memory, area, and performance overhead.

\section{Background}\label{sec:background}

\subsection{Hard Errors}
The impact of radiation on semiconductor devices can be categorized into soft and hard errors.
While soft errors are temporary and can often be corrected through normal operation or a power cycle, hard errors are permanent and result in irreversible damage~\cite{han2021single}.
The common causes of hard errors are total ionizing dose (TID), single-event burnout (SEB), and electrostatic discharge (ESD).
Such hard errors can be modeled as stuck-at-faults, bridge faults, or delay faults.
In extreme cases, hard errors cause short or open circuits on transistors connecting VDD to the GND, which makes the circuit nonoperational.

\subsection{Delay-based Sensors}
In glitching and radiation attacks, the injected energy causes disturbances on the power line of the chip.
Such disturbances can affect the propagation delays of electrical signals in delay-based sensors, such as ring oscillators (ROs)~\cite{gravellier2019high} and time-to-digital converters (TDCs)~\cite{zick2013sensing,zhao2018fpga}.
These sensors have been used to detect voltage, electromagnetic (EM), and laser glitching and radiating attacks on FPGAs, measuring voltage fluctuations on the power delivery network (PDN) caused by injected energy~\cite{monfared2024laserescape}.
The main drawbacks of using ROs are their high power consumption and sensitivity to environmental conditions, making them unattractive for space applications requiring low power consumption.
Therefore, in this paper, we deploy TDC sensors for attack detection. 
\begin{figure}[t]
\centering
\includegraphics[width=\columnwidth]{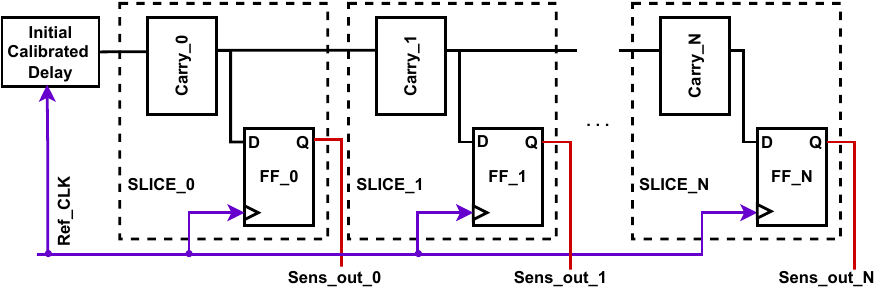}
\caption{high-level implementation of a TDC-based fault detection sensor}
\label{fig:tdc_sensor}
\end{figure}

We adopt the TDC implementation described by Mahmoud et.al.~\cite{mahmoud2023practical}, which is primarily designed for remote side-channel attacks on FPGA. We repurpose the TDC design for fault-sensing applications. 
A high-level illustration of the employed TDC is shown in Fig.~\ref{fig:tdc_sensor}. The fundamental building blocks of the sensor comprise an initial delayed signal followed by a tapped delay line and the corresponding output registers. The calibration is carried out offline using adjustable delay elements (e.g., IDELAY~\cite{IDELAY}) or by including/excluding combinational logic blocks in a chain to ensure that the sensor's output is in meta-stable status. 
Ideally, this means that for a \textit{N-bit} TDC sensor, calibration is achieved if output tends to a value with $HW(Output)=N/2$ during a long enough sampling time, where $HW$ represents the Hamming Weight.
The observable delay line is often deployed by fast carry propagation logic units constrained physically to organize a chain of delay. The output of each of these delay units is then propagated to an output register clocked with the original signal. Based on the propagation delay of the input of each register, the output can be constituted as a binary sensor.

\begin{figure*}[t] 
\centering
\includegraphics[width=0.85\linewidth]{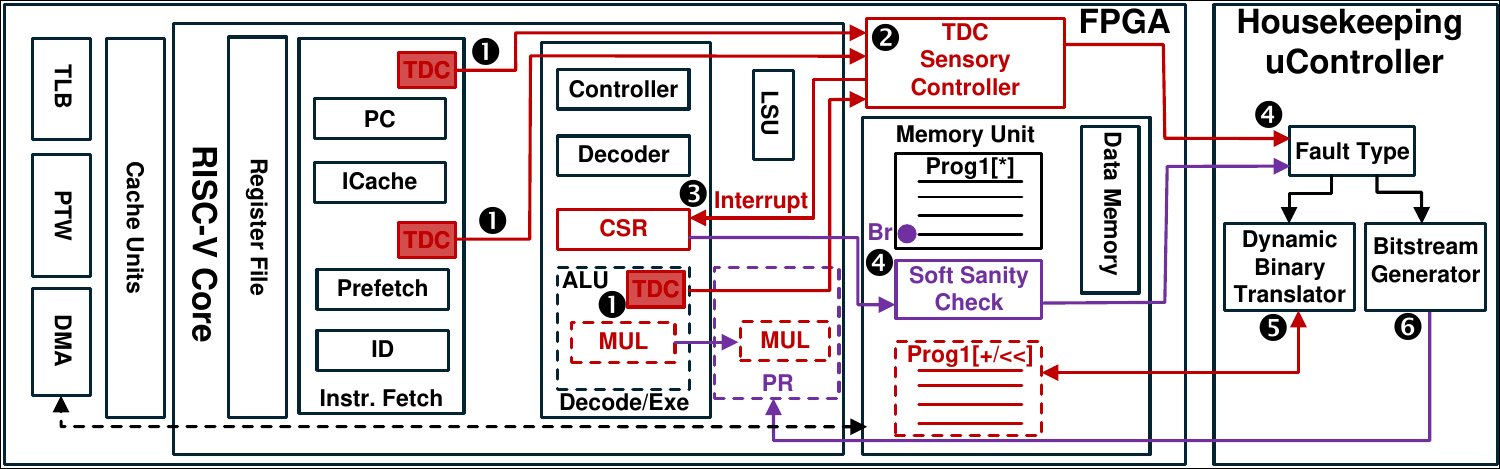}
\caption{The Proposed framework}
\label{fig:framework}
\end{figure*}

\subsection{Software Instruction Synthesis}
The arithmetic logic unit (ALU) is a fundamental building block of various computing machines.
The question here is if a computer can accomplish an arithmetic or logical task if one or more components (e.g., Multiplier, XOR, etc.) of the ALU fail.
To answer this question, we should understand to what extent it is possible to perform the failed operation/instruction with other available operations/instructions.

Various instruction set architectures (ISA) define how software communicates with the hardware.
While there are complex instruction set computers (CISC), such as x86 architecture, in which a single instruction can execute several low-level operations, reduced instruction set computers (RISC) have more simplified instructions in exchange for requiring more instructions to accomplish some complex tasks.
In principle, we can reduce the number of instructions to a single instruction to build a One-instruction set computer (OISC)~\cite{mavaddat1988urisc,nurnberg2004grand} or even a computer without instructions. 
For instance, it has been shown that the x86 {\tt mov} instruction~\cite{dolan2013mov}, x86 memory management unit (MMU) page faults~\cite{bangert2013page}, or Remote Direct Memory Access (RDMA) offloads are Turing-complete~\cite{reda2022rdma}, and hence, all instructions can be accomplished by one instruction or operation. 

While an OISC is an extreme architecture with a very large overhead in terms of execution time, it is still possible in the case of failure to substitute individual arithmetic or logical instructions by combining and synthesizing other instructions.
For example, if the processor's multiplier fails, the processor can still execute the multiplication using add and shift instructions.
Therefore, in our hard error scenario, depending on the location of an error, a processor with a partially damaged ALU might still be able to accomplish tasks though in a longer time.

\subsection{Hardware Partial Reconfiguration}
Partial Reconfiguration (PR) is a feature of mainstream field programmable gate arrays (FPGAs), allowing for dynamic modifications of a portion of the FPGA circuit without needing a complete system reboot. 
At the same time, the rest of the system continues to operate uninterrupted,~\cite{koch2012partial}.
The use of PR in FPGAs is particularly advantageous in mission-critical applications, like space applications~\cite{fuchs2018dynamic}, where system downtime is not acceptable.

AMD/Xilinx's Dynamic Function eXchange (DFX) introduces an approach for defining Partially Reconfigurable (PR) regions within the FPGA fabric\cite{xilixreconfig}.
However, the Vivado toolchain presents significant limitations, including slow performance for real-time applications and a lack of bitstream relocation support, constraining the reconfigurability~\cite{manev2022byteman}.
Several notable tools have emerged to address these challenges. 
For instance, the open-source tool Byteman~\cite{manev2022byteman} has substantially improved bitstream manipulation capabilities. 
It enhances efficiency, speed, and compatibility by supporting the merging of clock, Configurable Logic Block (CLB), and Block RAM data, along with implementing different merge strategies. These bitstream manipulation tools are crucial, as they provide the ability to generate and deploy partial bitstreams in realtime.
Such real-time partial reconfigurations have been recently deployed as countermeasures against side-channel attacks~\cite{monfared2024randohm,monfared2024laserescape}, where a processor can choose between multiple partial bitstreams generated in an offline phase and partially reconfigure the circuits on the programmable logic fabric to add noise to physical information leakages.
Hence, a housekeeping RadHard processor can use a similar method to replace and reroute defective components of a soft processor in orbit.

\section{Threat Model}
For our threat model, we assume two types of adversaries, namely, a cyber-physical adversary and a non-kinetic adversaries~\cite{swope2024space}.
We also consider cosmic radiation as an unintentional threat.
We assume that there is a RadHard housekeeping microcontroller onboard, which is resilient to radiations and glitches and can control commercial and non-rad-tolerant computer chips.
We only consider scenarios where the core of the non-rad-tolerant computer chips is partially damaged, but its communication links to the housekeeping MCU and its JTAG are still functional. Otherwise, our countermeasure will not be useful if the entire chip is fried or it cannot be programmed anymore.
Finally, we assume that FPGA scrubbing and ECC are deployed to detect and correct other faults, such as memory contents. 

\subsection{Cyber-physical Adversaries}
For the cyber-physical attackers, we assume a hosted payload environment in which a satellite bus accommodates multiple payloads by different vendors.
The satellite's bus resources, such as power lines, data buses, and physical space, are shared between the payloads.
We assume that one of the payloads is infected either by a remote cyber adversary or a supply-chain adversary using malware or hardware Trojans.
Therefore, the infected payload can perform physical faults injection attacks, such as voltage glitches/conducted interference through the shared power line or EM glitches/radiated interference through the shared unshielded space inside the bus enclosure, on the adjacent payload's computers or even the computer of the infected system itself, see Fig.~\ref{fig:threat}. 
The adversary's goal is to affect their functionality or reliability.
Examples of such remote fault attacks have been reported for multi-tenant FPGA~\cite{gnad2017voltage,alam2019ram} deployments in the cloud, where the adversary can cause soft and hard errors on the chip.

\subsection{Non-kinetic Adversaries}
For the non-kinetic adversaries, we assume that directed energy weapons (from a ground station or another satellite) or nuclear detonation in space are deployed to damage the satellite electronics in orbit.
While such attacks are not considered cyber threats, they still can partially damage the onboard electronics and computers, and their effect could be similar to cyber-physical attacks, though on a much larger scale.
Directed energy weapons, such as high-energy lasers, can thermally affect the satellite components and cause a current surge in the power/data lines of the onboard computer.
On the other hand, a nuclear blast in space generates high-energy electromagnetic pulses in the form of X-ray and Gamma radiation, which can ionize the material and create electrical currents that can damage computer chips.

\subsection{Cosmic Radiation}
Finally, the source of faults could be natural.
Cosmic rays are high-energy particles or particle clusters, mainly protons or atomic nuclei, traveling through space at nearly the speed of light. 
They originate from the Sun, our galaxy, or even distant galaxies. 
These high-energy particles can penetrate the enclosures and chip packages and permanently damage the transistors.

\section{Approach}\label{sec:approach}

This section describes our proposed approach to address hardware-level faults in real-time. Our solution utilizes a hardware/software co-design to mitigate hardware faults in the mission-critical processors deployed in space applications. Concurrently, we seek to incur minimal hardware modification to the existing hardware architectures to preserve the highest level of compatibility. We consider a soft processor on a reconfigurable fabric, such as FPGA, and exploit the reconfigurability to tackle localized permanent hardware damage.
Additionally, our main focus is the hardware permanent errors on the silicon as soft errors, such as bitflips in off-core memory units, can be corrected using scrubbing techniques~\cite{brosser2014assessing}, error correction codes~\cite{spensky2021glitching}, re-execution~\cite{de2010relax,ni2013acr}, redundancy methods~\cite{wirthlin2016seu}, and partial reconfiguration~\cite{fuchs2018dynamic}.

Figure \ref{fig:framework} depicts a high-level representation block diagram of the proposed framework. We consider a soft-core implementation of a RISC-V architecture. 
We equipped it with additional digital sensors distributed throughout the FPGA design. Highlighted in \circled{1}, TDC sensors are placed in different physical locations on the core and are calibrated to detect local disturbance during the execution. 
Furthermore, we incorporate a hardware sensor controller that constantly checks the output of the sensors.
The TDC controller in \circled{2} keeps track of environmental conditions for each building block of the processing core.
All sensors are tagged and sampled with a relatively higher sampling rate in comparison with the core's clock frequency. 
The controller verifies the condition of each local block by comparing the tagged sensor data with its pre-defined threshold that is derived during the calibration. 
Moreover, we extend the architecture to support an external non-maskable interrupt (NMI) from the sensor controller hardware. If a fault attack is initiated, the environmental disturbance changes the behavior of the TDC sensors, which is detected by the controller in real time. 
As a result, the core is interrupted with the highest priority of NMI~\cite{nmi} as illustrated in \circled{3}. 
Upon the occurrence of the interrupt, the processor breaks into a pre-defined software check where it verifies the functionality of its different components. 
As shown in \circled{4}, the sanity check software performs a comprehensive test and transmits the test's results to the off-chip Radiation Hardened \textit{Houskeeping $\mu$-controller}.
At the same time, based on the sensory data captured from the TDCs, tags for the potentially faulty building blocks are sent to the Housekeeping system. The fault type detected could be due to several failures diagnosed by the soft sanity check or a sensor abnormality report. Depending on the fault, a software-based response can be taken into account by utilizing a binary translator. As highlighted in \circled{5}, if the indicated fault type, for instance, is associated with \textit{Multiplier} part of the ALU, a real-time machine-code level translator is invoked to transform the target program to use non-faulty units (e.g., instructions). On the other hand, if multiple execution units are (or about to be ) permanently damaged, a full/partial reconfiguration command is sent from the $\mu$controller to the FPGA (e.g., Internal Configuration
Access Port ICAP~\cite{artix7} ) to re-implement \& reconfigure the affected units as indicated in \circled{6}. Note that the reconfiguration process incurs a relatively higher timing overhead than the binary translation method. However, in scenarios where permanent faults are manifested in multiple logic units, the system cannot be recovered by software, and reconfiguration is the only option for recovery. 

\begin{figure}[t]
\centering
\includegraphics[width=\columnwidth]{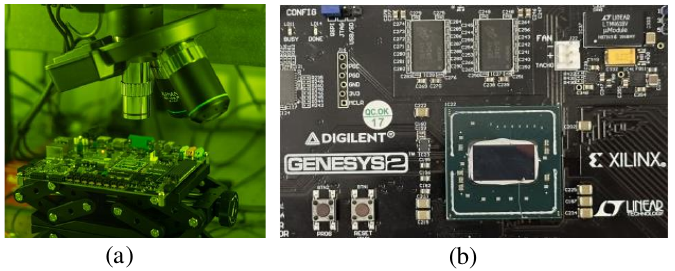}
\caption{(a) Device under test under the objective lenses of ALpHANOV setup. (b) The Kintex 7 FPGA chip is in a flip-chip package with a removed heatsink from its backside silicon.}
\label{fig:setup}
\end{figure}

\section{Experimental Setup}\label{sec:setup}

\subsection{Device Under Test}
We used two Genesys 2 development kits~\cite{genesys} for the experiments.
These kits contain an AMD/Xilinx Kintex 7 (XC7K325T-2FFG900C) FPGA manufactured with 28 nm technology.
The die of the FPGA under test is packaged in a flip-chip package, enabling 
access to the chip's backside silicon and exposing the FPGA fabric to laser radiation.
We did not perform any other modifications (e.g., silicon polishing or thinning) to the package or board.
The FPGA core was supplied by 1.0 V for all the experiments, and the global clock operated at 200 MHz.
One kit is used as a reference for all experiments, and the second one was intentionally damaged by the laser beam for hard error emulation. 

\subsection{Laser Setup}
We used an ALPhANOV S-LMS~\cite{Alphanov_2023} setup for photon emission microscopy, laser radiation, and fault injection. 
The microscope consists of a camera system for navigating, an InGaAs camera for photon mission analysis, and a 1064 nm wavelength laser source.
The microscope has been equipped with multiple lenses with 2.5X, 20X~(NA=0.6), and 50X~(NA=0.7) magnifications on an XYZ stage to focus on a region.
The combined setup is controlled using the software and hardware switches to control the XYZ stage and camera options. 
The software provides an IR view of the die, which can be used for navigation.
The peak currents used for radiation and hard error experiments were 1~A and 2.5~A, respectively, while the pulse width was 250~ns with a frequency of 100 kHz. 
The laser is controlled by the ALPhANOV control software in combination with viewing the die using the camera software. The combination results in a live feed to laser shots on the desired fault region (a screenshot shown in~\ref{fig:setup}).
The InGaAs camera was used to verify the FPGA's damage location by capturing the created short-circuits' photon emission. 

\subsection{Hardware Implementation}
\begin{figure}[t]
\centering
\includegraphics[width=\columnwidth]{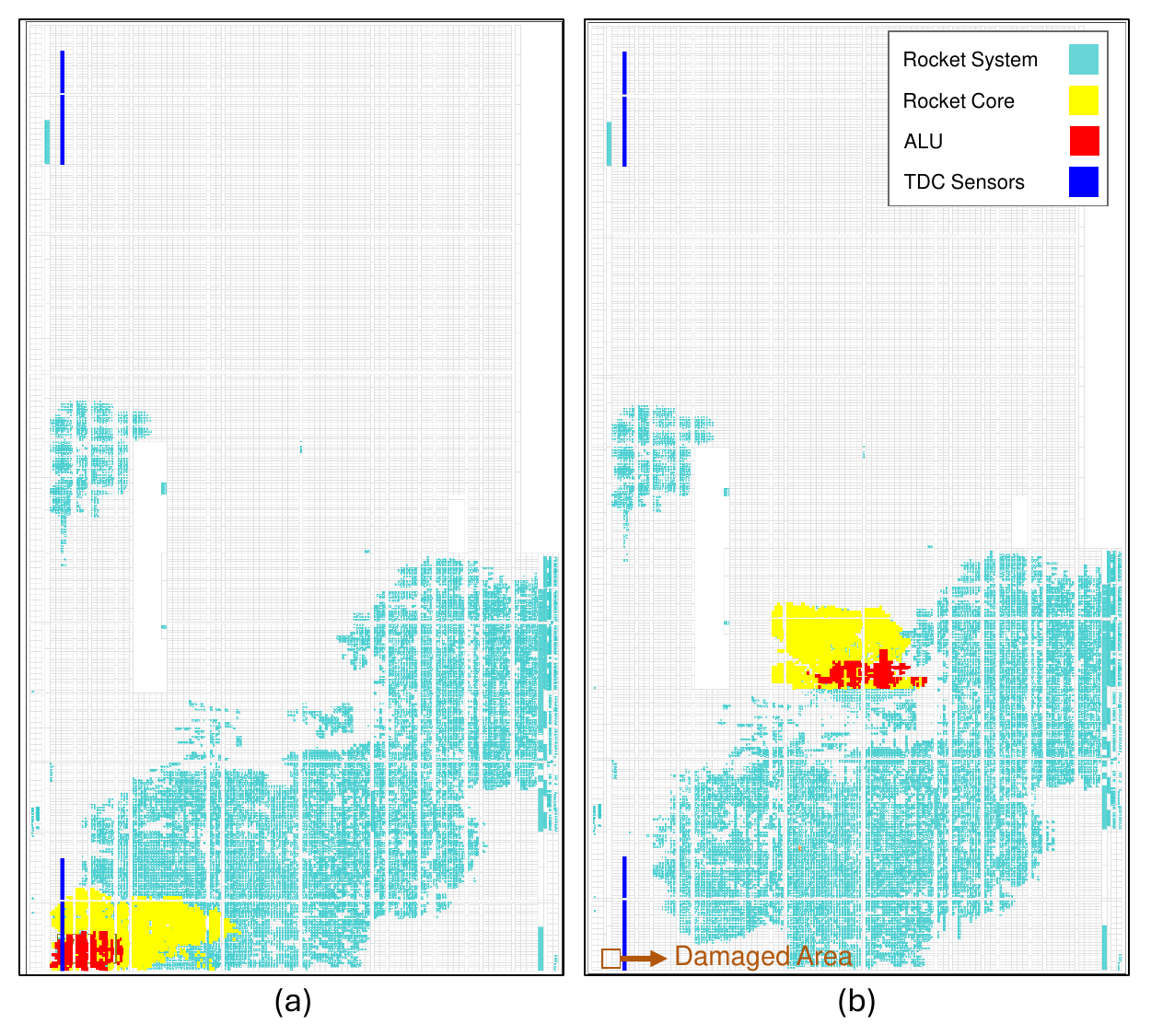}
\caption{Two different layouts of the entire Rocket System implementation on Kintex-7 FPGA. The region highlighted in yellow is the core excluding the ALU, and the region highlighted in red is the ALU. (a) The ALU and core are close to the TDC sensor. (b) The ALU and core are replaced and rerouted from the defect region.}
\label{fig:design}
\end{figure}

A single-core, 32-bit RISC-V is implemented on the FPGA under test. 
We used the open-source Rocket Chip generator to instantiate the RISC-V Rocket Core~\cite{asanovic2016rocket}.
We used an SD card to store and load the executable and linkable format (ELF) file containing bare-metal programs for the processor.
We used two parallel UARTs, one for communicating with the running code on the processor and the other one for reading back the sensors' values.
In our experiments, we used a computer to emulate the housekeeping microcontroller unit (MCU) to manage the software recompilations and FPGA's reconfigurations.
As illustrated in Fig.~\ref{fig:design} (a), two TDC sensors (represented by blue colors) are positioned in different parts of the FPGA. 
One of these sensors is placed near the ALU embedded within the processor to maximize the likelihood of detecting errors.
Table~\ref{table1} provides an overview of the FPGA resource utilization by various modules in the Rocket System in terms of the number of registers and Look-Up Tables (LUTs).
There is a hierarchical relationship between the three modules. 
The Rocket System module encompasses the complete system, including Rocket Core and ALU components.
The Rocket Core module contains only the processor core, including the ALU.
Peripheral and additional systems (e.g., JTAG and I/Os) are not part of this module. The TDC sensor modules have a minimal resource consumption, and thus, their overheads are negligible. For this implementation, we employed two TDC sensors, each with a width of 128 bits and  32 LUT\_5s for the calibration circuitry.
The LUT count for ALU reflects 19.4\% usage of Rocket Core, and most of its resources are consumed by the multiplier.


\vspace{15pt}
\begin{table}[h]
\noindent
\centering
\caption{Usage Resources}
\resizebox{\columnwidth-1cm}{!}{%
{\tiny
\begin{tabular}{ccc}
\hline
\textbf{Module} & \textbf{Registers} & \textbf{LUTs}   \\ \hline
Rocket System & \num{6350} & \num{15359} \\ \hline
Rocket Core   & \num{1557} & \num{3179}  \\ \hline
ALU           & \num{125}    & \num{617}    \\ \hline
TDC Sensors   &  \num{320}   & \num{64}  \\ \hline
\end{tabular}%
}
}
\label{table1}
\end{table}

\begin{figure*}[t]
\centering
\includegraphics[width=\textwidth]{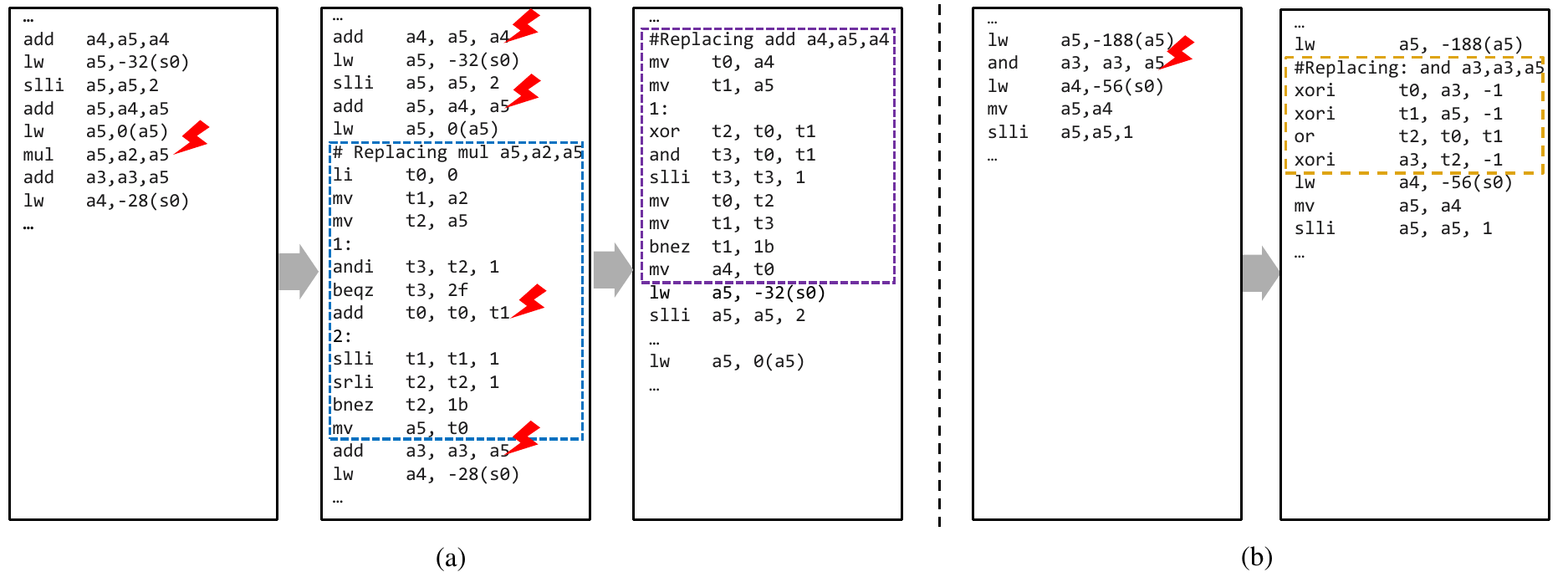}
\caption{(a) Translation and resynthesis of arithmetic instructions: Example of substitution of {\tt mul} instructions upon the failure of the multiplier. Further substitution of {\tt add} instruction in the case of addition failure. (b) Translation and resynthesis of logic instructions: Example of substitution of {\tt and} instructions upon the failure of the AND gate in ALU.}
\label{fig:resyntheis}
\end{figure*}

\subsection{Software Implementation}
We selected two benchmarks, namely, a multiply-accumulate (MAC) function along with AND operations and a Reed-Solomon coding scheme. 
Both benchmarks have been coded using the C programming language.
The C codes were compiled into assembly codes.
We wrote a translator script to automatically replace and substitute the instructions in the generated output assembly codes.
Using the translator, four variations of the assembly code for each benchmark were generated. 
In the first version, the original code is used without any modifications. 
In the second version, all multiplier operations are replaced with equivalent add-and-shift operations. 
In the third version, the multiplication and addition operations are substituted with XOR and AND functions. 
In the last version, to consider the effect of logic failure, the AND operation is replaced with other logic operations.

A script reads the outputs of the TDC sensors via UART. 
If an error is detected, a diagnostic test routine (soft sanity check) is executed to identify the faulty submodule. 
This routine determines whether the issue lies with the multiplier, adder, shift register, or AND components. At this stage, a command is sent to the FPGA via the UART interface to execute a different variation of the code based on the identified faulty module. 
\section{Results}\label{sec:results}

To evaluate our framework, we consider two fault scenarios that involve hard errors on the target's fabric. Specifically, first, we consider a temporary hardware fault induced by a laser beam and showcase how it is detected and mitigated by the proposed solution. Subsequently, we tackle the scenario where a physical part of the target chip is permanently damaged during the runtime. 
In the following, we distinctly detail each of these scenarios and describe our mitigation to address them.

\subsection{Radiation Sensing}
The radiation sensing system utilizes distributed digital sensors (i.e., TDC) throughout the target IC as described in  Sect.~\ref{sec:approach}. We evaluate radiation by statically analyzing digital samples extracted from the sensors and performing a comparative examination between normal and radiating environments emulated via NIR laser exposure.
\subsubsection{Normal Operation}
\begin{figure}[t]
\centering
\includegraphics[width=\columnwidth]{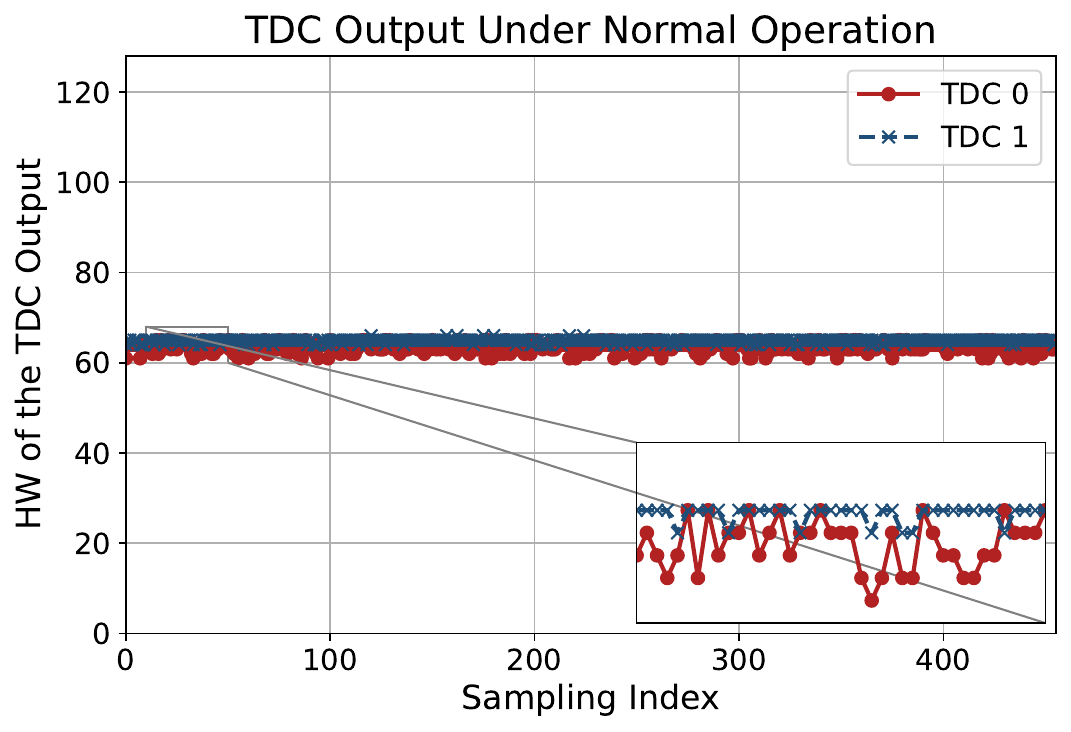}
\caption{The TDC outputs represented by HW during a normal operation of the target device}
\label{fig:tdc_plot_normal}
\end{figure}

During the \textit{Normal Operation}, the system is set to execute simple bare-metal applications. 
Simultaneously, two TDC sensors are deployed on the target's fabric as part of our solution in accordance with Fig.~\ref{fig:design}. As explained in Sect.~\ref{sec:approach}, the sensory controller keeps track of the output samples of these TDCs. For the sake of illustrious, Fig.~\ref{fig:tdc_plot_normal} demonstrates the samples captured by the controller unit in the system. As anticipated, the HW of the TDC samples tends to show a consistent trend fixed around the value of $128/2=64$, where 128 is the width of the sensor. During the sampling process, a slight alteration is observed specifically for \textit{TDC0}, which is physically located near the ALU circuitry. This alteration is possibly due to the local power consumption of the logical unit near the TDC0 sensor. In contrast, \textit{TDC1}'s output depicts a more consistent behavior, possibly due to the fact that it is deployed far away from the execution logic of the core (Please refer to Fig.~\ref{fig:design}).

\subsubsection{Temporary Laser Exposed Operation}

\begin{figure}[t]
\centering
\includegraphics[width=\columnwidth]{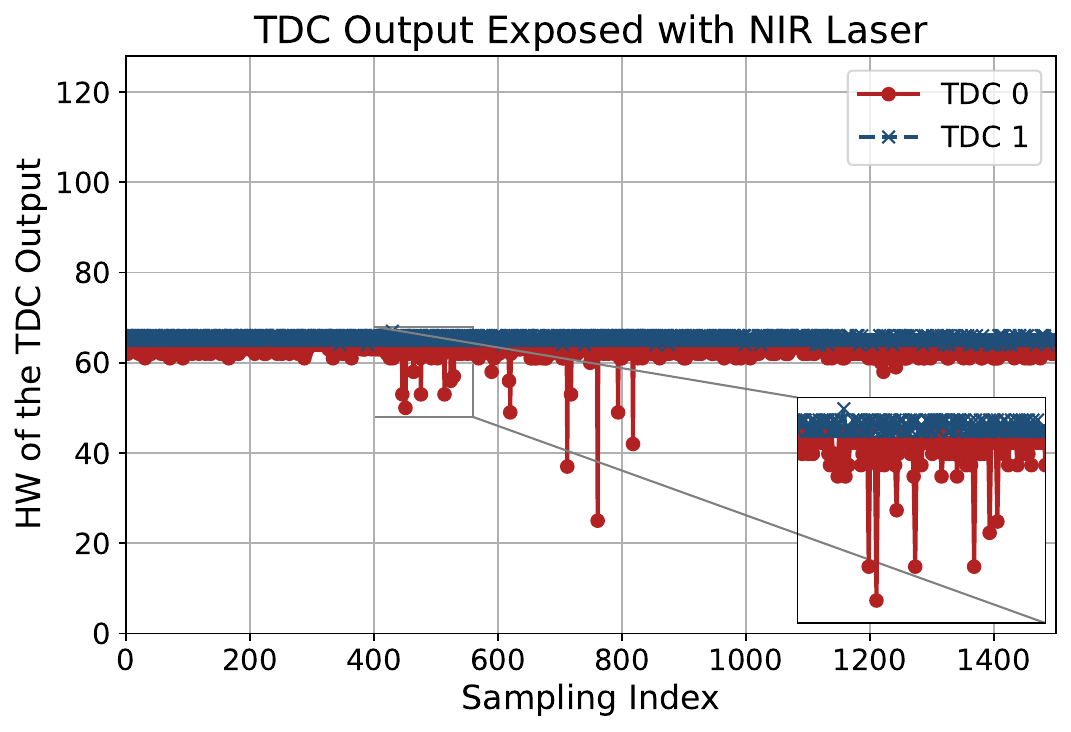}
\caption{The TDC outputs represented by HW when the target is exposed with NIR Laser}
\label{fig:tdc_plot_laser}
\end{figure}

Under the radiation emulation, we expose the backside of the IC with the laser with a setup as described in Sect.~\ref{sec:setup}.  We limit the peak current usage of the laser to 1.0 A to ensure that neither temporary nor permanent faults are induced in the target IC. Our experiment shows that an average peak current of 1.5 A is required to induce temporary faults (soft errors) on this particular fabric. As a result, here, we merely evaluate our system against radiations that do not cause a fault. This is specifically aligned with scenarios where the target device is about to be exposed to a vulnerable environment in orbit or radiation power is swept by the adversary.
In our experiments, we particularly 1) localized the ALU sub-system 2.5X lens, 2) carried out the focusing procedure with the 20X lens, and 3) activated the laser output and performed manual navigation for 5-10 seconds. During this experiment, we captured the TDC outputs from the sensor controller to evaluate the detection mechanism. Furthermore, note that as shown in Fig.~\ref{fig:design}, the \textit{TDC0} is physically deployed in the vicinity of the ALU. Fig.~\ref{fig:tdc_plot_laser} shows a sampling window of TDC's output during the laser exposure. As the laser stage is moved to expose blocks near the ALU, we can visually identify distinct jumps in the output of \textit{TDC0}, which is located near the target logic. In our framework (see Sect.~\ref{sec:approach}), this fluctuation indicates a potential fault attack and invokes the mitigation mechanism in software and hardware, which are outlined in the following.

\subsection{Instruction Resynthesis}
Upon detection of an error in the ALU using the soft sanity check (see Sect.~\ref{sec:approach}), the housekeeping microcontroller executes an algorithm to replace the affected assembly instructions.
Here, we present the results of two test scenarios.
In the first scenario, one or more arithmetic units, such as the multiplier and adder of the processor, fail.
In the second scenario, a logical unit, e.g., the AND gate, fails.
Fig.~\ref{fig:resyntheis}(a) shows the basic steps needed to substitute the {\tt mul} instruction with {\tt add} and shifting instructions~\cite{baruch2002structure}. 
The algorithm starts with the loading of the multiplicand number to the {\tt t1} register and a load of the multiplier number to the {\tt t2} register; it initializes the {\tt t0} register to 0 as a product register. 
The least significant bit of the multiplier register ({\tt t2}, calculated by {\tt and}ing it with 1) determines whether the multiplicand is added to the product register. Shifting the multiplicand to the left results in shifting of the intermediate products left.
The shift of the multiplier to the right, on the other hand, prepares the next bit of the multiplier to be tested in the next step. It is done in the loop until {\tt t2} is zero. 
If the adder of the processor also fails, the RISC-V {\tt add} instruction should be replaced with an implementation using bitwise operations for a ripple-carry adder, see Fig.~\ref{fig:resyntheis}(a).
The ripple-carry algorithm splits the addition into two separate operations: Computing the sum without carrying using the {\tt xor} operation and Computing the carry using the {\tt and} operation. Instead of handling all bits in parallel, we propagate the carry bit iteratively. The loop guarantees that no carry is left unprocessed, effectively completing the addition.


In the second scenario, where a logical operation like AND fails, the {\tt and} instruction can be replaced by using {\tt not} and {\tt or} instructions based on De Morgan's Laws:
\[
(A \land B) = \neg(\neg A \lor \neg B)
\]
According to the above equation, we can replace an AND operation by inverting the two operands, performing the OR operation on the inverted values, and then inverting the result. 
In the RISC-V assembly code, inversion is achieved by XORing a register with -1 (0xFFFFFFFF). 
The assembly language equivalent of an {\tt and} instruction requires the steps of Fig.~\ref{fig:resyntheis}(b).


\begin{figure}[t]
\centering
\includegraphics[width=\columnwidth]{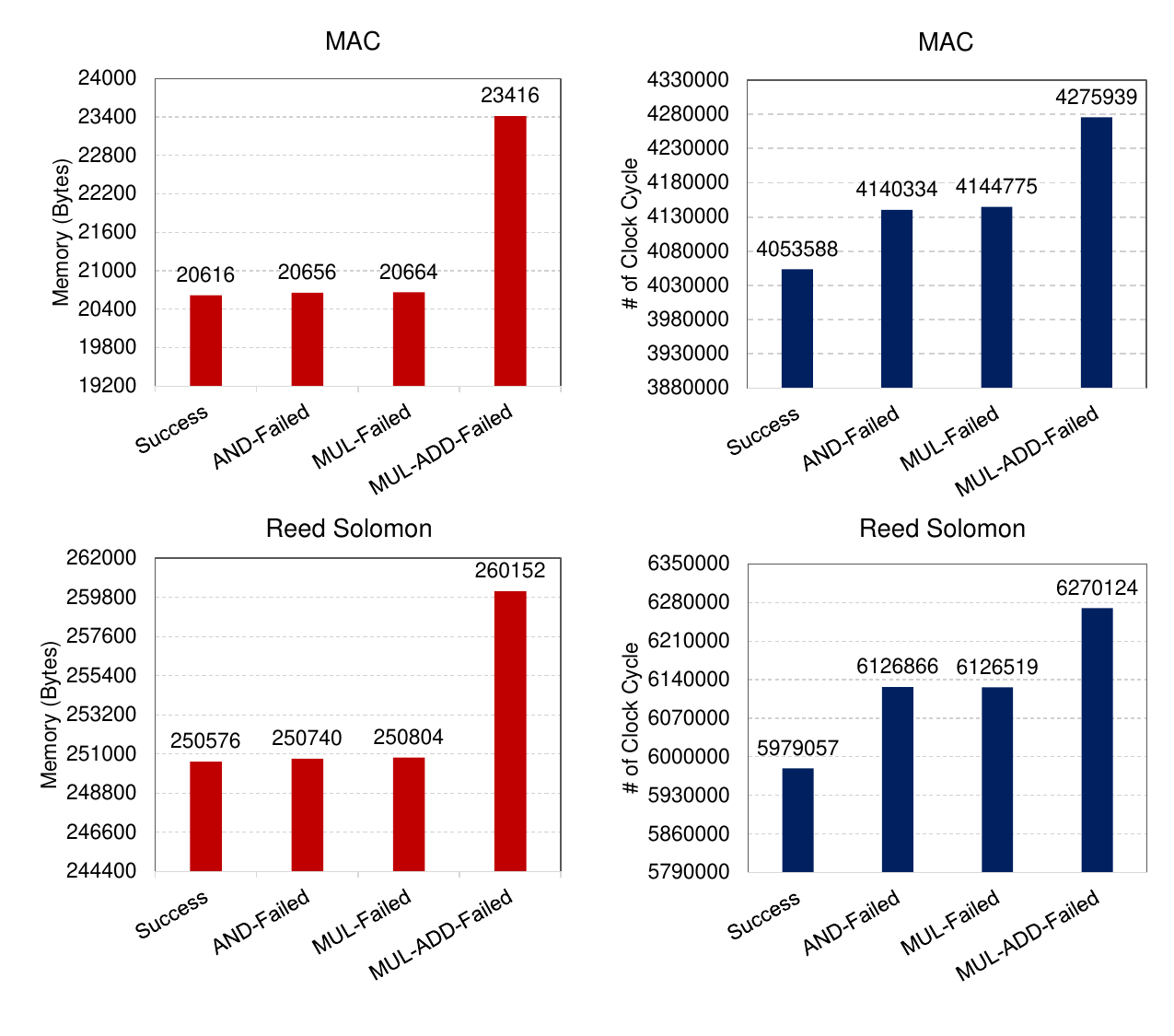}
\caption{The comparison of overhead in terms of memory usage and number of clock cycles in various scenarios in which different ALU operations fail.}
\label{fig:overhead}
\end{figure}

\begin{figure*}[t]
\centering
\includegraphics[width=0.9\textwidth]{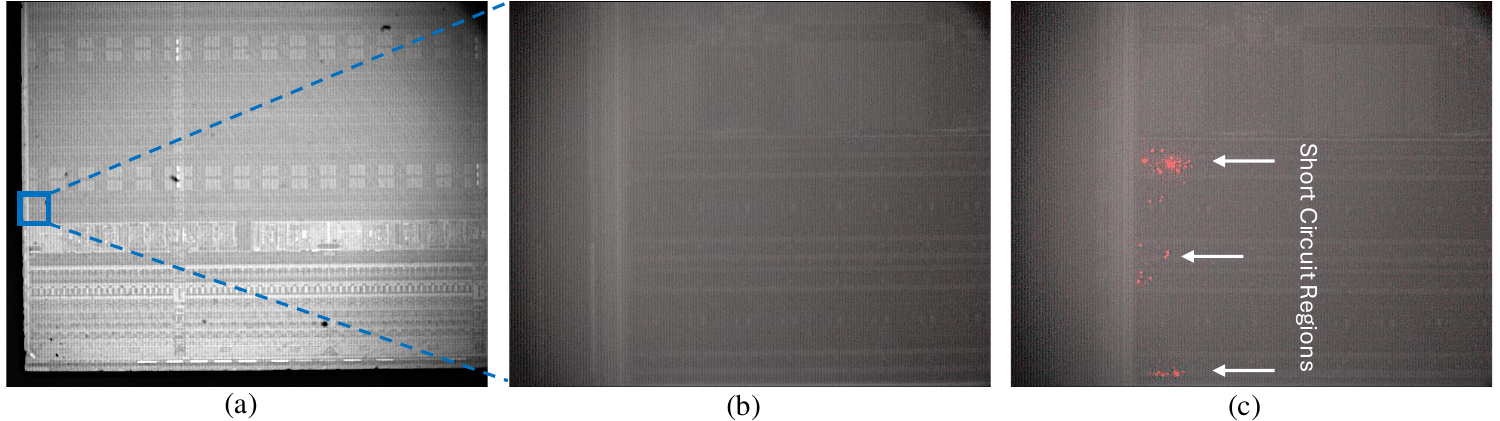}
\caption{(a) The reflectance image of one corner of the FPGA with a highlighted region of the injected fault. (b) The zoomed-in photon emission of healthy LUTs of the undamaged FPGA. (c) The zoomed-in photon emission of short-circuited LUTs of the damaged FPGA.}
\label{fig:emission}
\end{figure*}

Fig.~\ref{fig:overhead} illustrates the overhead in calculations in terms of memory and time resources. 
Specifically, the memory usage of MAC and Reed-Solomon benchmarks demonstrates that {\tt and} and {\tt mul} failures result in an increase of a few bytes of additional memory compared to the original code. 
In contrast, the failure of {\tt add} and {\tt mul} together require 2800 and 9576 bytes of additional memory in MAC and Reed Solomon, respectively.
This significant difference is due to the increased number of {\tt add} instructions in the assembly code.
As expected, increasing the number of instructions also affects the number of clock cycles. 
As illustrated in Fig.~\ref{fig:overhead}, both the {\tt mul} failure and the {\tt and} failure take almost the same amount of time to complete the total operation. 
However, while the memory requirements for these two scenarios are comparable to the memory usage of the original code, the time required to execute these tasks is substantially larger. 
This is because {\tt mul} instruction is being replaced with loop-based instructions. 
Furthermore, the time required for execution of the recovery code when both {\tt add} and {\tt mul} have failed together is longer than for all other forms of the MAC function and Reed-Solomon due to the increased number of {\tt add} instructions. 
However, because a single addition operation takes less time than a single multiplication, the time ratio between function variations does not directly correspond to memory utilization. 
Note that we only considered the time overhead imposed on instruction execution.
The latency overhead associated with the binary translator or communication with the MCU depends on the onboard computer architecture and deserves a separate study.

\subsection{Hardware Reconfiguration for Extensive HW Damage}
As explored in the previous section, if radiation causes faults to the limited number of blocks of ALU, the computation can be detected and consequently recovered with software resynthesis. 
However, if extensive hardware damage occurs due to a powerful fault injection attack, where, for instance, the routing of the signals is permanently affected, a hardware recovery is necessary. 
Here, we present a significantly low-overhead reconfiguration technique to recover from major permanent hardware failures compared to existing core redundancy approaches.
Fig.~\ref{fig:emission} demonstrates the images captured from the FPGA. In Fig.~\ref{fig:emission}(a), a zoomed-out layout of the FPGA is depicted with a 2.5X lens. The highlighted blue box indicates the coordinates where the ALU is implemented in accordance with Fig.~\ref{fig:design}(a). 
We set the core to perform its normal operation. To emulate a wide-spread hard fault and evaluate our mitigation, we increase the peak current utilization of our NIR laser up to 2.5~A and perform a laser injection focused on the LUTs, realizing the ALU blocks on the IC. As the exposure of the laser is initiated, the core immediately crashes and fails to provide the anticipated output. On the other hand, we observed that the TDC sensory controller indicates a possible fault in our setup. We also observed that the permanent faults caused by the laser resulted in local short circuits on the FPGA. Fig.~\ref{fig:emission}(c) shows a high photon emission activity due to the short circuits in the damaged FPGA after the high-energy laser exposure. Fig.~\ref{fig:emission}(b) depicts the same physical layout in an undamaged FPGA. Under normal circumstances, the damaged FPGA could not be utilized as an operational processor. However, based on the TDC sensory data captured right before the laser exposure, our framework is able to perform a PR and re-implement the entire core physically positioned near the TDCs in an undamaged location. By considering the reconfiguration time overhead of AMD/Xilinx's ICAP, we verified the correctness of the core and successful recovery of the system in the new layout shown in Fig.~\ref{fig:design}(b).
Note that reserved LUTs/FFs are required for hardware reconfiguration. 
The number of reserved LUTs/FFs must match the number of resources needed for the Rocket Core and ALU logic.
Moreover, depending on the application and the availability of LUTs, TDCs can be duplicated across all locations in the FPGA. Such duplication assists in more precise localized fault detection while imposing negligible area overhead. 


\section{Conclusion}\label{sec:conclusion}
In this paper, we presented a detection- and response-based countermeasure for processor chips in space applications to mitigate hard errors caused by adversarial and natural faults.
We showed that FPGA-compatible TDC sensors can detect radiation and physical glitching attempts on the processor chip's PDN and issue a warning for sanity checking.
Upon detection of a fault, depending on the level of corruption, software resynthesis for substituting the corrupted instructions or hardware reconfiguration for relocating/rerouting the processor circuit will be carried out to recover the functionality.
To assess the effectiveness of our approach, we performed laser radiation and fault injection on a soft RISC-V processor realized on an FPGA.
Our results showed that the processor can continue its operation even with the presence of hard errors (e.g., short circuits) to accomplish the arithmetic and logical tasks.

%
\IEEEpeerreviewmaketitle

\section*{Acknowledgment}
This effort was sponsored by NSF Grant CNS-2150123.




%



\bibliographystyle{IEEEtran}
\bibliography{references.bib}

\begin{thebibliography}{10}
\providecommand{\url}[1]{#1}
\csname url@samestyle\endcsname
\providecommand{\newblock}{\relax}
\providecommand{\bibinfo}[2]{#2}
\providecommand{\BIBentrySTDinterwordspacing}{\spaceskip=0pt\relax}
\providecommand{\BIBentryALTinterwordstretchfactor}{4}
\providecommand{\BIBentryALTinterwordspacing}{\spaceskip=\fontdimen2\font plus
\BIBentryALTinterwordstretchfactor\fontdimen3\font minus \fontdimen4\font\relax}
\providecommand{\BIBforeignlanguage}[2]{{%
\expandafter\ifx\csname l@#1\endcsname\relax
\typeout{** WARNING: IEEEtran.bst: No hyphenation pattern has been}%
\typeout{** loaded for the language `#1'. Using the pattern for}%
\typeout{** the default language instead.}%
\else
\language=\csname l@#1\endcsname
\fi
#2}}
\providecommand{\BIBdecl}{\relax}
\BIBdecl

\bibitem{bailey2021cybersecurity}
B.~Bailey, ``Cybersecurity protections for spacecraft: A threat based approach,'' \emph{The Aerospace Corporation}, 2021.

\bibitem{l3harris}
L3Harriss, ``{Hosted Payload Interface Unit (HPIU)},'' \url{[Online] https://www.l3harris.com/all-capabilities/hosted-payload-interface-unit-hpiu}, 2019.

\bibitem{warning}
W.~Hennigan, ``{The Warning},'' \emph{The New York Times}, 2024.

\bibitem{markoff2022tiny}
J.~Markoff, ``Tiny chips, big headaches.'' \emph{International New York Times}, pp. NA--NA, 2022.

\bibitem{dixit2021silent}
H.~D. Dixit, S.~Pendharkar, M.~Beadon, C.~Mason, T.~Chakravarthy, B.~Muthiah, and S.~Sankar, ``Silent data corruptions at scale,'' \emph{arXiv preprint arXiv:2102.11245}, 2021.

\bibitem{hochschild2021cores}
P.~H. Hochschild, P.~Turner, J.~C. Mogul, R.~Govindaraju, P.~Ranganathan, D.~E. Culler, and A.~Vahdat, ``Cores that don't count,'' in \emph{Proceedings of the Workshop on Hot Topics in Operating Systems}, 2021, pp. 9--16.

\bibitem{moini2023fault}
S.~Moini, D.~Kansagara, D.~Holcomb, and R.~Tessier, ``Fault recovery from multi-tenant fpga voltage attacks,'' in \emph{Proceedings of the Great Lakes Symposium on VLSI 2023}, 2023, pp. 557--562.

\bibitem{paquette2021visualizing}
M.~Paquette, B.~Marquis, R.~Bainbridge, and J.~Chapman, ``Visualizing electromagnetic fault injection with timing sensors,'' in \emph{2021 IEEE Physical Assurance and Inspection of Electronics (PAINE)}.\hskip 1em plus 0.5em minus 0.4em\relax IEEE, 2021, pp. 1--8.

\bibitem{monfared2024laserescape}
S.~K. Monfared, K.~Mitard, A.~Cannon, D.~Forte, and S.~Tajik, ``{LaserEscape: Detecting and Mitigating Optical Probing Attacks},'' in \emph{2023 IEEE/ACM International Conference on Computer Aided Design (ICCAD)}, 2024.

\bibitem{yuce2018secure}
B.~Yuce, C.~Deshpande, M.~Ghodrati, A.~Bendre, L.~Nazhandali, and P.~Schaumont, ``A secure exception mode for fault-attack-resistant processing,'' \emph{IEEE Transactions on Dependable and Secure Computing}, vol.~16, no.~3, pp. 388--401, 2018.

\bibitem{jerosecuring}
S.~Jero, J.~Furgala, M.~A. Heller, B.~Nahill, S.~Mergendahl, and R.~Skowyra, ``{Securing the Satellite Software Stack},'' in \emph{SpaceSec}, 2024.

\bibitem{alam2019ram}
M.~M. Alam, S.~Tajik, F.~Ganji, M.~Tehranipoor, and D.~Forte, ``Ram-jam: Remote temperature and voltage fault attack on fpgas using memory collisions,'' in \emph{2019 Workshop on Fault Diagnosis and Tolerance in Cryptography (FDTC)}.\hskip 1em plus 0.5em minus 0.4em\relax IEEE, 2019, pp. 48--55.

\bibitem{han2021single}
J.-W. Han, M.~Meyyappan, and J.~Kim, ``Single event hard error due to terrestrial radiation,'' in \emph{2021 IEEE International Reliability Physics Symposium (IRPS)}.\hskip 1em plus 0.5em minus 0.4em\relax IEEE, 2021, pp. 1--6.

\bibitem{gravellier2019high}
J.~Gravellier, J.-M. Dutertre, Y.~Teglia, and P.~Loubet-Moundi, ``High-speed ring oscillator based sensors for remote side-channel attacks on fpgas,'' in \emph{2019 International conference on ReConFigurable computing and FPGAs (ReConFig)}.\hskip 1em plus 0.5em minus 0.4em\relax IEEE, 2019, pp. 1--8.

\bibitem{zick2013sensing}
K.~M. Zick, M.~Srivastav, W.~Zhang, and M.~French, ``Sensing nanosecond-scale voltage attacks and natural transients in fpgas,'' in \emph{Proceedings of the ACM/SIGDA international symposium on Field programmable gate arrays}, 2013, pp. 101--104.

\bibitem{zhao2018fpga}
M.~Zhao and G.~E. Suh, ``Fpga-based remote power side-channel attacks,'' in \emph{2018 IEEE Symposium on Security and Privacy (SP)}.\hskip 1em plus 0.5em minus 0.4em\relax IEEE, 2018, pp. 229--244.

\bibitem{mahmoud2023practical}
D.~G. Mahmoud, O.~Glamo{\v{c}}anin, F.~Regazzoni, and M.~Stojilovi{\'c}, ``Practical implementations of remote power side-channel and fault-injection attacks on multitenant fpgas,'' in \emph{Security of FPGA-Accelerated Cloud Computing Environments}.\hskip 1em plus 0.5em minus 0.4em\relax Springer, 2023, pp. 101--135.

\bibitem{IDELAY}
Xilinx. (2017) Zynq-7000 soc (z-7007s, z-7012s, z-7014s, z-7010, z-7015, and z7020): Dc and ac switching characteristics data sheet(ds187). \url{https: //docs.xilinx.com/v/u/en-US/ds187-XC7Z010-XC7Z020-Data-Sheet}.

\bibitem{mavaddat1988urisc}
F.~Mavaddat and B.~Parhami, ``Urisc: the ultimate reduced instruction set computer,'' \emph{International Journal of Electrical Engineering Education}, vol.~25, no.~4, pp. 327--334, 1988.

\bibitem{nurnberg2004grand}
P.~J. N{\"u}rnberg, U.~K. Wiil, and D.~L. Hicks, ``A grand unified theory for structural computing,'' in \emph{Metainformatics: International Symposium, MIS 2003, Graz, Austria, September 17-20, 2003. Revised Papers}.\hskip 1em plus 0.5em minus 0.4em\relax Springer, 2004, pp. 1--16.

\bibitem{dolan2013mov}
\BIBentryALTinterwordspacing
S.~Dolan, ``{{\tt mov} is Turing-complete},'' 2013. [Online]. Available: \url{https://drwho.virtadpt.net/files/mov.pdf}
\BIBentrySTDinterwordspacing

\bibitem{bangert2013page}
J.~Bangert, S.~Bratus, R.~Shapiro, and S.~W. Smith, ``The $\{$Page-Fault$\}$ weird machine: Lessons in instruction-less computation,'' in \emph{7th USENIX Workshop on Offensive Technologies (WOOT 13)}, 2013.

\bibitem{reda2022rdma}
W.~Reda, M.~Canini, D.~Kosti{\'c}, and S.~Peter, ``$\{$RDMA$\}$ is turing complete, we just did not know it yet!'' in \emph{19th USENIX Symposium on Networked Systems Design and Implementation (NSDI 22)}, 2022, pp. 71--85.

\bibitem{koch2012partial}
D.~Koch, \emph{Partial reconfiguration on FPGAs: architectures, tools and applications}.\hskip 1em plus 0.5em minus 0.4em\relax Springer Science \& Business Media, 2012, vol. 153.

\bibitem{fuchs2018dynamic}
C.~M. Fuchs, N.~M. Murillo, A.~Plaat, E.~van~der Kouwe, and T.~P. Stefanov, ``Dynamic fault tolerance through resource pooling,'' in \emph{2018 NASA/ESA Conference on Adaptive Hardware and Systems (AHS)}.\hskip 1em plus 0.5em minus 0.4em\relax IEEE, 2018, pp. 9--16.

\bibitem{xilixreconfig}
Xilinx. (2023, Dec) Xilinx introduction to dynamic function exchange. \url{https://docs.xilinx.com/r/en-US/ug909-vivado-partial-reconfiguration/Introduction-to-Dynamic-Function-eXchange}.

\bibitem{manev2022byteman}
K.~Manev, J.~Powell, K.~Matas, and D.~Koch, ``byteman: A bitstream manipulation framework,'' in \emph{2022 International Conference on Field-Programmable Technology (ICFPT)}.\hskip 1em plus 0.5em minus 0.4em\relax IEEE, 2022, pp. 1--9.

\bibitem{monfared2024randohm}
S.~K. Monfared, D.~Forte, and S.~Tajik, ``Randohm: Mitigating impedance side-channel attacks using randomized circuit configurations,'' in \emph{2023 IEEE/ACM International Conference on Computer Aided Design (ICCAD)}, 2024.

\bibitem{swope2024space}
C.~Swope, K.~A. Bingen, M.~Young, M.~Chang, S.~Songer, and J.~Tammelleo, ``Space threat assessment 2024,'' 2024.

\bibitem{gnad2017voltage}
D.~R. Gnad, F.~Oboril, and M.~B. Tahoori, ``Voltage drop-based fault attacks on fpgas using valid bitstreams,'' in \emph{2017 27th International Conference on Field Programmable Logic and Applications (FPL)}.\hskip 1em plus 0.5em minus 0.4em\relax IEEE, 2017, pp. 1--7.

\bibitem{brosser2014assessing}
F.~Brosser, E.~Milh, V.~Geijer, and P.~Larsson-Edefors, ``Assessing scrubbing techniques for xilinx sram-based fpgas in space applications,'' in \emph{2014 International Conference on Field-Programmable Technology (FPT)}.\hskip 1em plus 0.5em minus 0.4em\relax IEEE, 2014, pp. 296--299.

\bibitem{spensky2021glitching}
C.~Spensky, A.~Machiry, N.~Burow, H.~Okhravi, R.~Housley, Z.~Gu, H.~Jamjoom, C.~Kruegel, and G.~Vigna, ``Glitching demystified: analyzing control-flow-based glitching attacks and defenses,'' in \emph{2021 51st Annual IEEE/IFIP International Conference on Dependable Systems and Networks (DSN)}.\hskip 1em plus 0.5em minus 0.4em\relax IEEE, 2021, pp. 400--412.

\bibitem{de2010relax}
M.~De~Kruijf, S.~Nomura, and K.~Sankaralingam, ``Relax: An architectural framework for software recovery of hardware faults,'' \emph{ACM SIGARCH Computer Architecture News}, vol.~38, no.~3, pp. 497--508, 2010.

\bibitem{ni2013acr}
X.~Ni, E.~Meneses, N.~Jain, and L.~V. Kal{\'e}, ``Acr: Automatic checkpoint/restart for soft and hard error protection,'' in \emph{Proceedings of the international conference on high performance computing, networking, storage and analysis}, 2013, pp. 1--12.

\bibitem{wirthlin2016seu}
M.~J. Wirthlin, A.~M. Keller, C.~McCloskey, P.~Ridd, D.~Lee, and J.~Draper, ``Seu mitigation and validation of the leon3 soft processor using triple modular redundancy for space processing,'' in \emph{Proceedings of the 2016 ACM/SIGDA International Symposium on Field-Programmable Gate Arrays}, 2016, pp. 205--214.

\bibitem{nmi}
``Ibex risc-v reference guide: Exceptions and interrupts,'' \url{https://ibex-core.readthedocs.io/en/latest/03\_reference/exception\_interrupts.html}, accessed: 2024-12-10.

\bibitem{artix7}
Xilinx. (2023, May) Xilinx 7 series fpgas configurable logic block. \url{https://www.eng.auburn.edu/~nelson/courses/elec4200/FPGA/ug4747SeriesCLB.pdf}.

\bibitem{genesys}
Digilent, ``{Genesys 2},'' \url{[Online] https://digilent.com/reference/programmable-logic/genesys-2/start}, 2023.

\bibitem{Alphanov_2023}
AlphaNov, ``{Single Laser Fault Injection Microscope - S-LMS},'' \url{[Online] https://www.alphanov.com/en/products-services/single-laser-fault-injection}, 2024.

\bibitem{asanovic2016rocket}
K.~Asanovic, R.~Avizienis, J.~Bachrach, S.~Beamer, D.~Biancolin, C.~Celio, H.~Cook, D.~Dabbelt, J.~Hauser, A.~Izraelevitz \emph{et~al.}, ``The rocket chip generator,'' \emph{EECS Department, University of California, Berkeley, Tech. Rep. UCB/EECS-2016-17}, vol.~4, pp. 6--2, 2016.

\bibitem{baruch2002structure}
Z.~F. Baruch, \emph{Structure of computer systems}.\hskip 1em plus 0.5em minus 0.4em\relax UT Pres, 2002.

\end{thebibliography}

\end{document}